\documentclass[aps,prl,twocolumn]{revtex4}
\RequirePackage{xspace}
\usepackage{graphicx}

\def\be{\begin{equation}}
\def\ee{\end{equation}}
\def\bearr{\begin{eqnarray}}
\def\eearr{\end{eqnarray}}
\def\hlf{${\frac{1}{2}}$~}
\begin{document}
\preprint{IMSc-2004/10/XX}

\title{Possibility of magnetic field induced Composite Fermi sea in neutral Graphene}

\author{ G. Baskaran\email}
\affiliation{%
Institute of Mathematical Sciences, C.I.T. Campus, Chennai 600 113, India }%

\begin{abstract}
Neutral graphene in strong magnetic fields is believed to be an (exchange stabilized)  integer Hall state of completely filled up spin (say) and empty down spin bands of n = 0, two fold valley degenerate Landau levels. We suggest that correlation energy gain from a nearly SU(4) singlet (2 spin $\times$ 2 valley) composite fermi sea formation, at $\nu \approx$ \hlf filling for each component, destabilizes ferromagnetic integer quantum Hall state. This radically different scenario is consistent with a dissipative gapless state seen in experiments in neutral graphene. Interesting paired Hall states are possible, from residual interactions, in this SU(4) fermi sea with a small Zeeman spin polarization.
\end{abstract}
\maketitle

A flurry of theoretical and experimental activities in graphene has followed discovery of a successful way\cite{prep} to cleave atomically thin graphene layers, make electrical contacts and perform measurements. The efforts have been rewarding, as graphene exhibits some surprising electrical properties, in addition to some unusual quantum Hall behavior\cite{Novoselov05,Zhang05,Novoselov04,Zhang06,Abanin07}. Many of these works focus on the behavior of charged graphene, controlled by external gates. However, we lack a deeper understanding of the normal state of neutral graphene in the presence of strong magnetic field, even before we charge it. 

Recent experimental\cite{Zhang06,Abanin07} and theoretical\cite{Nomura06,Abanin06a} works suggest a ferromagnetic integer Hall state at n = 0, where an exchange (and to a lesser extent Zeeman) split two (valley) degenerate levels are completely filled (say) with upspin electrons (fig 1A). There is a puzzle however: even though experiments see a weak Hall plateau beyond about 20 T, diagonal resistance $\rho_{xx}$ is very high and metal like, instead of being zero. It seem to indicate presence of a gapless dissipative state. We propose a resolution to the puzzle (fig 1B) by suggesting a composite fermi sea normal state for neutral graphene. Our proposal, if confirmed, has important consequences, as composite fermi seas are one of the strangest form of quantum matter. For example, composite fermi seas are seats of paired Hall states and non Abelian quasi particles. Non Abelian quasi particles are currently being vigorously studied in the context of topological quantum computation\cite{dasSarmaPhysToday06}.

There are many theoretical studies in graphene in the presence of strong external magnetic and electric fields, dealing with both integer and fractional quantum Hall states\cite{gusyninIQHE,Abanin06a,Fertig06,Nomura06,Alicea06,Goerbig06,Apalkov06,herbut,Gusynin94,Khveshchenko01,Gusynin06,Fuchs06,vinu1}. A recent experimental study \cite{Zhang06} probes nature of low temperature state of the neutral graphene in magnetic fields upto 45 T. Development of a weak Hall plateau at n = 0 is attributed to a ferromagnetic  integer quantum Hall state (fig 1A). However, they find an unexpected, dissipative normal state in neutral graphene. 
Another work confirms this\cite{Abanin07} and also suggests that the dissipative normal state is intrinsic and not a consequence of disorder. Thus a puzzle is the appearance of a very weak quantum Hall plateau and a contrasting (compressible, gapless) metallic and high resistive state; longitudinal resistance is finite and large, $\rho_{xx} > \frac{e^2}{h}$. This work attributes the dissipative state to certain anomalous behaviour of edge states, arising from a theoretically predicted `spin gap' behaviour\cite{Abanin06a}. Other, general theoretical study of fractional quantum Hall states in graphene, \cite{CFothers,DimoCF} discuss spin polarized composite fermi sea, but do not address the important experimentally motivated case being discussed in the present paper.
\begin{figure}
\includegraphics*[width=6.0cm]{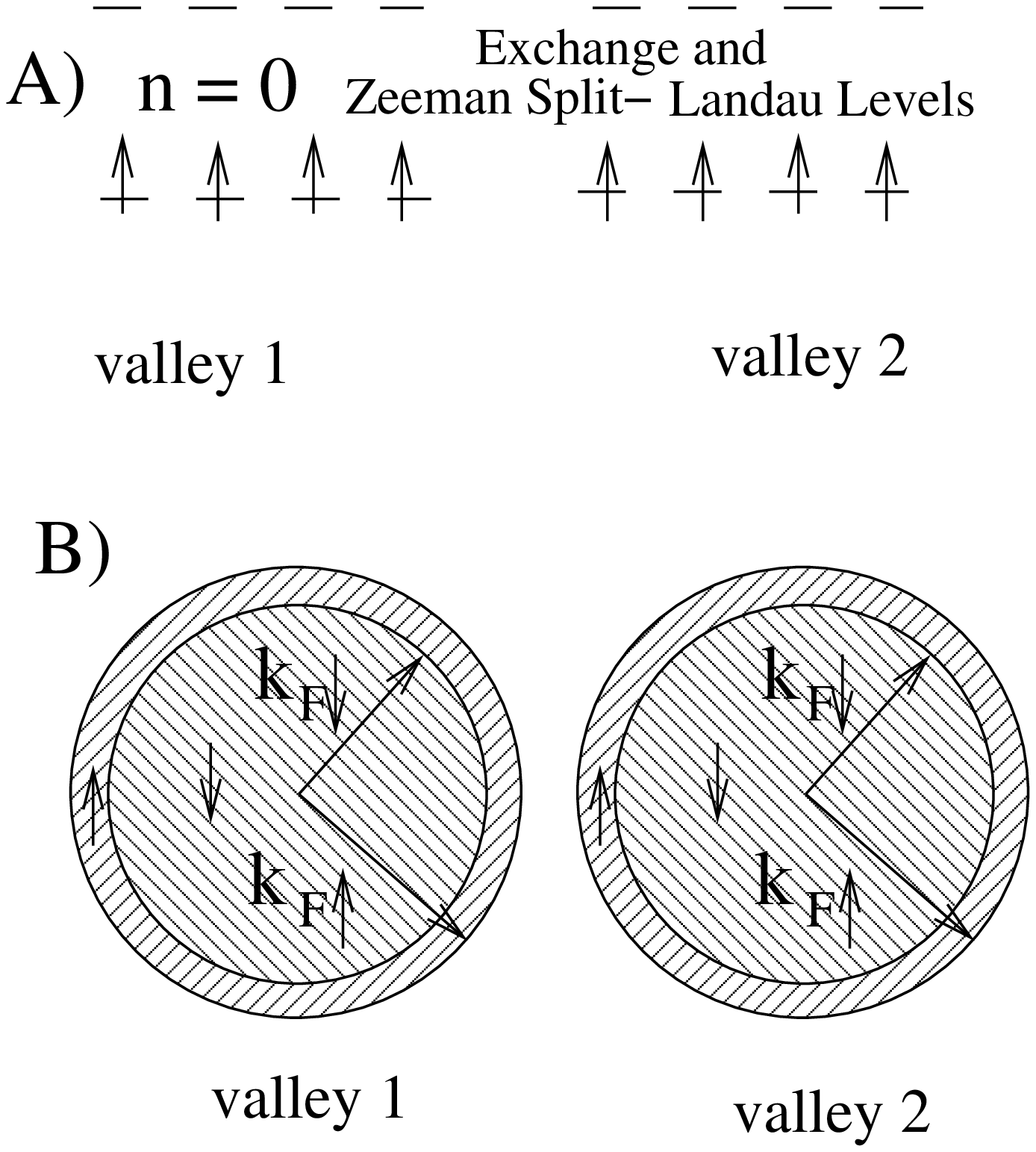}
\caption{\label{fig1}
A) Exchange and Zeeman split n = 0 two valley degenerate Landau levels and completely filled upspin bands.
B) Our proposal of an approximate SU(4) singlet (2 spin $\times$ 2 valley) composite fermi sea, at a filling $\nu \approx$ \hlf for each component. A small spin polarization (k$_{\rm F \uparrow} \neq$  k$_{\rm F \downarrow}$) arising from Zeeman field is shown.}
\end{figure}

The aim of the present paper is to suggest a rather different scenario, that is inspired by the aforementioned experimental results (see however, the note \cite{note}) on neutral graphene. We suggest that instead of the expected fully spin polarized integer quantum Hall state at n = 0, we have a SU(4) singlet (2 spin $\times$ 2 valley) composite fermi liquid, with each species at $\nu \approx$ \hlf filling. We present physical arguments for emergence of a composite fermi sea and a very rough estimate of energy of a variational composite fermi sea wave function. 

At the heart of our proposal are the following key physical arguments: a) Two (valley) n = 0 completely filled bands with parallel spins, lack inter band dynamical scattering because of complete Pauli blocking. (We ignore high energy n $\neq$ 0 Landau levels). Consequently two electrons from different valleys with parallel spins do not have a correlation hole; to that extent coulomb repulsion energy is not minimised. Further, the \textit{overlap charges} of two single particle states p and q, at any lattice site $\bf R$, $\psi^{*n=0}_{p,\tau}({\bf R})\psi^{n=0}_{q,\mu}({\bf R}) = 0$ for n = 0 Landau level states of different valleys $\tau$ and $\mu$ in graphene.\textit{ Thus there is no exchange energy between any two parallel spins belonging to different valleys ! }

b) Our composite fermi sea, a global SU(4) singlet (2 spin $\times$ 2 valley), on the other hand, is more efficient in building coulomb hole between electrons belonging to different valleys and also different spins (because of the half filled band character and less Pauli blocking), through two body scattering. The process of building correlation hole and avoiding coulomb repulsion will introduce short range SU(4) singlet correlation in the composite fermi sea. To this extent \textit{our proposal is somewhat similar to generic stability of spin liquid or antiferromagnetic state, compared to  ferromagnetic state in half filled band of strongly correlated electrons}.

Further, for graphene,  correlation and exchange energies are both comparable in magnitude and $ \sim \frac{e^2}{\epsilon_0 \ell_B} \approx 130 {\sqrt B} K$ (B measured in Tesla); Zeeman energy is negligible for B $\sim $ 30 T. In order to get a good ground state both correlation and exchange energies need to be considered.

We focuss on the n = 0 Landau level of our neutral graphene and assume that the lattice parameter of graphene `a' is small compared to the magnetic length$\ell_B$; i.e. $\frac{a}{\ell_B} << 1$. Appropriately normalized lattice coordinates of an electron are complex numbers $z_{\sigma\tau}$, where $\sigma = \uparrow,\downarrow$ and $\tau = \pm$ are the spin and valley indeces. 

We will make a brief remark about Jain's composite fermion\cite{jainCF,jainBook,HLR} approach. In 2D quantum Hall problems, external magnetic fields generate quantized vortices. Stable many body states are formed, when there is a commensurate relation between total number of vortices and total number of electrons. In most of the stable quantum Hall states an electron gets `bound' (associated) to a finite number of vortices (flux quanta) to become a quasi particle, in terms of which the complex many body problem becomes essentially non-interacting. When number of bound vortices are even, statistics of the composite object remains a fermion. They are the composite fermions. As these composites have already absorbed the effect of external magnetic field they see an effective magnetic field $B^* < B $, the externally applied field. At the end of some hierarchies (eg. $p \rightarrow \infty,~~ \frac{p}{2p + 1}$ = \hlf), the effective magnetic field seen by a composite fermion vanishes, leading to a compressible composite fermi liquid state. Composite fermion formation is a profound modification of the bare constituent electron. It also truns out to be an efficient way to build correlation/exchange holes and avoid coulomb repulsion.  

Now we will estimate and compare energies of spin polarized (upspin) completely filled n = 0 Landau levels of 2 valleys $\Psi_{\rm G}^{\uparrow\uparrow}$, with our proposed composite fermi sea state, $\Psi_{\rm G}^{\rm FS}$. The explicit form of the spin polarized state is
\be
\Psi_{\rm G}^{\uparrow\uparrow}[z] = \prod_{i<j,\tau}(z_{i\uparrow\tau} - z_{j\uparrow\tau}) e^{-\sum_{i\tau} \frac{|z_{i\uparrow\tau}|^2}{4}}
\ee
It is a product of two (antisymmetric) Van der Monde determinant for up spin electrons filling the n = 0 Landau levels of two vallyes,  $\tau = \pm$. Here $N_e \equiv 2N_\phi$ is the total number of electrons, and $4 N_\phi$ is the total number of single particle states in n = 0 Landau level, including spin and orbital degeneracies. And N$_\phi \approx
A \frac{eB}{hc}$, where A is the area of the sample.

In quantum Hall effect kinetic energy is frozen and is independent of electron-electron interaction. So one focuses only on two body electrostatic interaction energy. It is customary to write it as an energy gain with reference to electrostatic interaction energy of uniformly smeared electron charges and positve charges of neutralizing background. Energy of interaction of an electron with quantum numbers $\uparrow\tau$, with rest of the electrons is
\be
U_{\rm FM} = \pi \sum_{\tau'}\int \frac{e^2}{\epsilon_0 r} (g^{\uparrow\uparrow}_{\tau\tau'}(r) - 1) r dr
\ee
Here $g^{\sigma\sigma'}_{\tau\tau'}$(r) is the radial distribution function between two electrons having quantum numbers $\sigma\tau$ and $\sigma'\tau'$. For large distances 
g(r) $\rightarrow$ 1; so there is no energy gain. For short distances, comparable to $\ell_B$ and less, there is a `hole' in g(r) and a consequent energy gain by avoiding the strong  coulomb repulsions. One of the challenges in quantum Hall effect is to discover the right many body variation wave functions (built from appropriate single particle Landau level states) and get the best g(r).
The hole in g(r) in the present ferromagnetic state arises from antisymmetrization of electrons with parallel spins within a given valley. It is an exchange rather than correlation hole.

Within the lowest Landau level approximation $g^{\uparrow\uparrow}_{\tau -\tau}$(r) = 1; that is, there is no hole ! It indicates total absence of correlations between electrons belonging to two different valleys. This is a consequence of complete Pauli blocking of interband two body scattering among two filled bands. Consequently, short distance coulomb repulsion among electrons from two different valleys can not be avoided. Virtual scattering involving high energy n $\neq$ 0 Landau levels will to some extent modify  $g^{\uparrow\uparrow}_{\tau - \tau}$(r), which we ignore as a first approximation. 

As mentioned earlier, there is no exchange energy between two parallel spins belonging to different valleys in the n = 0 Landau level, as the overlap charge density between any two single particle states corresponding to different valleys are identically zero. 

Using Laughlin's plasma analogy, the expression\cite{laughlin83} for energy per electron is found to be
\be
U_{\rm FM} \approx \pi \int \frac{e^2}{\epsilon_0 r} (g^{\uparrow\uparrow}_{\tau\tau}(r) - 1) r dr \approx - 0.62 \frac{e^2}{\epsilon_0 \ell_B}
\ee

We will argue that we can get a lower energy in our composite fermi sea.
The wave function for the composite fermi sea we are proposing is given by
\bearr
\Psi_{\rm G}^{\rm FS}[z] &=&
\hat{P}_{LLL}  e^{-\sum_{i<j\sigma\tau\sigma'\tau'} 
f(|\textbf{r}_{i\sigma\tau} - \textbf{r}_{j\sigma'\tau'}|)}\nonumber \\  
&~&\prod_{\sigma\tau} 
\left\| e^{i \textbf{k}\cdot \textbf{r}_{\sigma\tau}}\right\|\prod_{i<j}(z_{i\sigma\tau} - z_{j\sigma\tau})^2e^{-\sum_i \frac{|z_{i\sigma\tau}|^2}{4}}
\eearr
There are 4 composite fermi sea of electrons, corresponding to two spin and two valley
indices. Each composite fermi sea has a filling fraction $\nu = \frac{1}{2}$ containing 
$\frac{N_\phi}{2}$ electrons. There is a common fermi momentum for each fermi sea given
by the expression: $k_F \equiv \sqrt{4\pi \rho}$. Here $\rho$ is electron density.

We have also introduced a two particle short range Jastrow factor through the function $f(r)$ between any two electrons. This introduces
correlation between two electrons having different valley and spin quantum numbers.
Since we have only a half filled Landau level valley exchange and spin exchange scattering within n = 0 levels are not completely Pauli blocked and they build short distance holes in the respective g(r), reduce the coulomb repulsion energy. 

Our short range Jastrow correlation for electrons with different quantum numbers  is unusual in quantum Hall situation. Long ranged Laughlin-Jastrow correlation are normally used, as they also take care of antisymmetry automatically. To keep a gapless fermi sea and at the same time maintain half filling in each Landau sub band, short range Jastrow function seems essential in the present case. When we attempt to use Laughlin-Jastrow factor, it either changes the mean density (expands the Laughlin drop significantly) or we seem to get incompressible states. Since one implements lowest Landau level projection at the end, in principle our short range Jastrow factor is allowed. We have no proof that our short range f(r) maintains the mean density per Landau level to be half. 

In the above variational wave function, the projection to n = 0 Landau level is done by $\hat{P}_{LLL}$. This projector effectively
replaces $\bar z \rightarrow \partial_z $. For example,
$e^{i \textbf{k}\cdot \textbf{r}} = e^{\frac{1}{2}i( \bar k z + k\bar z )}
\rightarrow e^{\frac{1}{2}i(\bar k z + k\partial_{\bar z})}$ and
$f(r_{ij}) \equiv 
\sum_k \tilde{f}(k) e^{i \textbf{k}\cdot \textbf{r}_{ij}}
\rightarrow  \sum_k \tilde{f}(k) e^{\frac{1}{2}i(\bar k z_{ij} + k\partial_{\bar z_{ij}})}$.
Here $k \equiv k_x + ik_y$.

Our composite fermi sea wave function is a global SU(4) singlet (spin $\times$ valley). That is, a plane wave state is either filled with four electrons with four possible spin and valley quantum numbers or is completely empty. In view of this it satisfies Fock's cyclic boundary condition. The above SU(4) singlet state is consistent with a two body interaction which is very nearly SU(4) symmetric for graphene in the n = 0 Landau level\cite{Goerbig06}.

As before, interaction energy of an electron with quantum numbers $\sigma\tau$
with rest of the electrons, in the composite fermi state is given by
\be
U_{\rm CF} = \pi \sum_{\sigma'\tau'}\int \frac{e^2}{\epsilon_0 r} 
(g_{\sigma \sigma'}^{\tau\tau'}(r) - 1) r dr
\ee

We will estimate the energy of this state in two approximations made on 
$|\Psi_{\rm G}^{\rm FS}[z]|^2$.  As the first approximation we put f(r) = 0.
We are left with products of four independent composite fermi sea, each with a filling
$\nu$ = \hlf. The energy of this state has been calculated in the literature\cite{jainCF,HLR} for spin polarized $\nu$ = \hlf composite fermi state as
\be
U_{\rm CF} \approx - 0.46 \frac{e^2}{\epsilon_0 \ell_B}
\ee
This energy is about 25 percent higher than the ferromagnetic integer Hall state (equation 3). We will show approximately that inclusion of a short range Jastrow correlation can give more than 25 percent energy gain and make our composite fermi sea stable. Jastrow factor, by construction is capable of generating hole in the radial distribution function $g^{\sigma\sigma'}_{\tau\tau'}(r)$. Physically this possibility arises from possibility of virtual interband scattering between electrons having different $\sigma\tau$ quantum numbers within the n = 0 Landau levels. Fortunately these scattering process are not completely Pauli blocked, because each Landau sub band is only half filled.

Without making a detailed calculation, we can estimate that a 25 percent improvement of energy is possible, by the following crude argument. From two particle scattering point of view, we have both direct and exchange scattering among two electrons with parallel spins. The corresponding matrix elements are comparable, and decay in a Gaussian fashion for localized gaussian orbitals separated beyond magnetic length $\ell_B$. For particles with different quantum numbers the exchange term is absent. That is, for our interband scattering, exchange scattering, half of the total processes are absent. Further, the interband scattering can take place to empty single particle states which are only half of the total number of states available (the rest are Pauli blocked). In view of this we approximate the amount of hole development in g(r) and the corresponding energy gain to be about $\frac{1}{4}$ = \hlf $\times$ \hlf of energy gain within a given composite fermi sea.
Since a given electron can reduce coulomb energy by correlating with three other composite fermi sea, the net energy we gain from the  3 $\times \frac{1}{4}$. Putting these numbers we
get our estimate of our state as:

\be
U_{\rm CF} \approx - (1 + \frac{3}{4}) ~ 0.46 \frac{e^2}{\epsilon_0 \ell_B} \approx
- 0.8 \frac{e^2}{\epsilon_0 \ell_B}
\ee 

This energy is lower than that of fully spin polarized filled Landau level states
(equation 3). It is the valley degeneracy which helps us to get a lower energy for the composite fermi sea !

We test our simple argument against a known case, namely stability of the standard n = 1 fully polarized quantum Hall state (no valley degeneracy) with  a spin singlet compsite fermi sea. Experimentally it is known (at least for lower odd integer Hall states) that fully polarized quantum Hall state always wins. Interestingly, in our estimate, absence of valley degeneracy reduces the factor $\frac{3}{4}$ to $\frac{1}{4}$ and keeps the spin polarized filled n = 1 state marginally stable.

After the composite fermi sea is formed, the Zeeman energy creates spin polarization
in an otherwise spin singlet composite fermi sea. The spin polarization of the composite fermi sea is easily estimated to be $\approx \frac{g\mu_B B}{U_{\rm CF}}$. We estimate that for graphene for a field of 30 T, spin polarization is less than a few percent. 

Very good signatures of composite fermion and fermi sea effects, in standard quantum Hall systems have been experimentally studied \cite{CFexp}. It will be very interesting to perform such studies and look for composite fermi sea in neutral graphene in strong magnetic fields. 

Composite fermions are neutral and they carry certain dipole moment\cite{reed}, as a function of their momenta. The response of the composite fermion fermi surface to external perturbations such as a local defect will be interesting. We will have a Friedel oscillation in dipole density. 

Further graphene may offer alternate methods to study composite fermions, because of new access through ARPES, STM etc, which are not possible in standard quantum Hall devices.

So far we have been talking about low temperature normal state. Residual interactions will introduce low temperature pairing instabilities in graphene either in the particle-particle or particle-hole channels. The small spin polarization will interfere with the standard instabilities. 

I wish to thank R. Shankar for bringing to my attention the dissipative character of the n = 0 quantum Hall state of graphene in strong magnetic field; R. Shankar and Vinu Lukose for critical comments.


\begin{references}
\bibitem{prep} K.S. Novoselov et al., Science, {\bf 306}, 666(2004)

\bibitem{Novoselov05}
K. S. Novoselov, {\it et al.}, Nature {\bf 438}, 197 (2005)

\bibitem{Zhang05}
Y. Zhang,
% {\it et al.},
Y.-W. Tan, H. L. Stormer and P. Kim, 
Nature {\bf 438}, 201 (2005) 

\bibitem{Novoselov04}
K. S. Novoselov, {\it et al.},Science, {\bf 306}, 666 (2004);
Proc. Natl. Acad. Sci. USA, {\bf 102}, 10451 (2005)

\bibitem{Zhang06}
Y. Zhang,
{\it et al.}, Phys. Rev. Lett., {\bf 96}, 136806 (2006)

\bibitem{Abanin07}
D. A. Abanin et al.,
cond-mat/0702125

\bibitem{dasSarmaPhysToday06} S. D. Sarma, M. Freedman, and C. Nayak, Physics Today 4, 123 (2006)

\bibitem{gusyninIQHE}V.P. Gusynin, S.G. Sharapov, Phys.Rev.Lett. {\bf 95}
146801(2005);C.L. Kane, E.J. Mele, Phys. Rev. Lett. {\bf 95} 146802(2005);
A. H. Castro Neto, F. Guinea, N. M. R. Peres, Phys.Rev.B {\bf 73}, 205408 (2006)

\bibitem{Abanin06a}
D. A. Abanin, P. A. Lee and L. S. Levitov, 
Phys. Rev. Lett. {\bf 96}, 176803 (2006);
cond-mat/0611062

\bibitem{Nomura06}
% Quantum Hall Ferromagnetism in Graphene
K. Nomura and A. H. MacDonald, Phys. Rev. Lett. {\bf 96}, 256602 (2006)
K. Yang, Kun; Das Sarma, A.H. MacDonald, Phys. Rev. {\bf B 74} 075423 (2006)


\bibitem{Fertig06}
H. A. Fertig and L. Brey, 
% Luttinger Liquid at the Edge of a Graphene Vacuum
% cond-mat/0604260, unpublished.
Phys. Rev. Lett. {\bf 97}, 116805 (2006); A. Iyengar et al.,
cond-mat/0608364 

\bibitem{Alicea06}
J. Alicea and M. P. A. Fisher, 
% Graphene integer quantum Hall effect in the ferromagnetic and paramagnetic regimes
Phys. Rev. B {\bf 74}, 075422 (2006)

\bibitem{Goerbig06}
M. O. Goerbig, R. Moessner, B. Doucot, Phys. Rev. B{\bf 74} 161407 (2006)

\bibitem{Apalkov06}
V. M. Apalkov, T. Chakraborty, 
Phys. Rev. Lett. {\bf 97}, 126801 (2006) 

\bibitem{herbut}I.F. Herbut, Phys. Rev. Lett., {\bf97} 146401 (2006)

\bibitem{Gusynin94}
V. P. Gusynin, V. A. Miransky, and I. A. Shovkovy, 
Phys. Rev. Lett. {\bf 73}, 3499 (1994)

\bibitem{Khveshchenko01}
D. V. Khveshchenko, Phys. Rev. Lett. {\bf 87}, 206401 (2001)
ibid. {\bf 87}, 246802 (2001)

\bibitem{Gusynin06}
V. P. Gusynin,
% {\it et al.},
V. A. Miransky, S. G. Sharapov, I. A. Shovkovy, 
% Excitonic gap, phase transition, and quantum Hall effect in graphene
% cond-mat/0605348, unpublished.
Phys. Rev. B{\bf 74}, 195429 (2006)

\bibitem{Fuchs06}
J.-N. Fuchs and P. Lederer, 
% cond-mat/0607480, unpublished.
Phys. Rev. Lett. {\bf 98}, 016803 (2007)

\bibitem{vinu1}
Vinu Lukose, R. Shankar and G. Baskaran (to appear in Phys. Rev. Lett.)
cond-mat/0603594

\bibitem{CFothers}
C. Toke et al., cond-mat/0606461; C. Toke and J. Jain, cond-mat/0701026.

\bibitem{DimoCF} D. V. Khveshchenko, cond-mat/0607174 


\bibitem{note}I thank R. Shankar and Vinu Lukose who point out that $\rho_{xx}$ and $\rho_{xy}$ seen in current experiments in graphene are quantitatively and some what qualitatively different from the well studied and clean $\nu =$ \hlf composite fermi sea systems (eg. H.W. Jiang et al., Phys. Rev. {\bf B40} 12013 (1989)) This could be due to a (nearly 200 times) lower mobility of graphene in present experiments. Further, a weak Hall plateau that develops in graphene, beyond about 20 T may indicate a first order phase transition of the normal state from low magnetic field composite fermi sea to a high field ferromagnetic state.

\bibitem{jainCF}
J. K. Jain, Phys. Rev. Lett. {\bf 63} 199 (1989);
Phys. Rev. {\bf B 41} 7653 (1990); Science {\bf 266} 1199 (1994)

\bibitem{jainBook}J.K. Jain, Composite Fermion (Cambridge University Press, Cambridge) 2007

\bibitem{HLR} B. I. Halperin, P. A. Lee, and N. Read, Phys. Rev. {\bf B 47} 7312 (1993);
E. H. Rezayi and N. Read, Phys. Rev. Lett. {\bf 72} 900 (1994) 

\bibitem{laughlin83} R. B. Laughlin, Phys. Rev. Lett. {\bf 50} 1395 (1983) 


\bibitem{CFexp}W. Kang, H.L. Stormer, L.N. Pfeiffer, K. Baldwin and K.W. West. Phys. Rev. Lett. {\bf 71}  (1993;
R.L. Willett, R.R. Ruel, K.W. West and L.N. Pfeiffer. Phys. Rev. Lett. {\bf71} (1993)
V.J. Goldman, B. Su and J.K. Jain. Phys. Rev. Lett. {\bf72} (1994)

\bibitem{reed}
N. Read, Semicond. Sci. Technol. {\bf 9}  1859 (1994);
Surf. Sci. {\bf 361/362} 7 (1996); G. Baskaran, Physica {\bf B 212}320 (1995)
\end{references}
\end{document}